\UseRawInputEncoding
\documentclass[twoside]{article}
\usepackage{amsmath,caption,threeparttable}
\usepackage{eurosym,mathrsfs,fancyhdr,CJK,multicol,graphics,indentfirst,bm,upgreek,booktabs,graphicx,multirow,warpcol}
\usepackage{epstopdf}
\usepackage{cite}
\usepackage{url}
\usepackage{float}

\usepackage{adjustbox}
\usepackage{array}
\usepackage{bbding}   % for checkmarks
\usepackage[edges]{forest}
\usepackage{wrapfig}
\usepackage{pdflscape}
\usepackage{latexsym}
\usepackage{amssymb}

\def\footnoterule{\kern 1mm \hrule width 10cm \kern 2mm}

\allowdisplaybreaks
\catcode`@=11
\def\title#1{\vspace{3mm}\begin{flushleft}\vglue-.1cm\Large\bf\boldmath\protect\baselineskip=18pt plus.2pt minus.1pt #1
\end{flushleft}\vspace{1mm} }
\def\author#1{\begin{flushleft}\normalsize #1\end{flushleft}\vspace*{-4pt} \vspace{3mm}}
\def\address#1#2{\begin{flushleft}\vglue-.35cm${}^{#1}$\small\it #2\vglue-.35cm\end{flushleft}\vspace{-2mm}\par}
\def\jz#1#2{{$^{\footnotesize\textcircled{\tiny #1}}$\let\thefootnote\relax\footnotetext{\!\!$^{\footnotesize\textcircled{\tiny #1}}$#2}}}
\catcode`@=11
\def\section{\@startsection{section}{1}{\z@}%
 {-3ex \@plus -.3ex \@minus -.2ex}%
 {2.2ex \@plus.2ex}%
{\normalfont\normalsize\protect\baselineskip=14.5pt plus.2pt minus.2pt\bfseries}}
\def\subsection{\@startsection{subsection}{2}{\z@}%
 {-3ex\@plus -.2ex \@minus -.2ex}%
 {2ex \@plus.2ex}%
{\normalfont\normalsize\protect\baselineskip=12.5pt plus.2pt minus.2pt\bfseries}}
\def\subsubsection{\@startsection{subsubsection}{3}{\z@}%
 {-2.2ex\@plus -.21ex \@minus -.2ex}%
 {1.4ex \@plus.2ex}
{\normalfont\normalsize\protect\baselineskip=12pt plus.2pt minus.2pt\sl}}

\begin{document}
\begin{CJK*}{GBK}{song}
\thispagestyle{empty}
\vspace*{-13mm}
\noindent {\small Pre-Print}
%===========================================================
\vspace*{2mm}

\title{A Survey on Evaluating Large Language Models in Code Generation Tasks}

\author{Liguo Chen$^{1}$, Qi Guo$^{1}$, Hongrui Jia$^{1}$, Zhengran Zeng$^{1}$, Xin Wang$^{1}$, Yijiang Xu$^{1}$, Jian Wu$^{3}$, Yidong Wang$^{1}$,Qing Gao$^{1}$, Jindong Wang$^{2}$, Wei Ye$^{1}$, Shikun Zhang$^{1}$$^{*}$}
\address{1}{Peking University, Beijing, China}
\address{2}{Microsoft Research Asia, Beijing, China}
\address{3}{Tokyo Institute of Technology , Tokyo, Japan}
% \address{5}{Affiliation of Author Three, Department, University, City, Country}

\let\thefootnote\relax\footnotetext{{}\\[-4mm]\indent\ Pre-Print}
\let\thefootnote\relax\footnotetext{{}\\[-4mm]\indent\ * corresponding authors.}
% \footnotetext[1]{corresponding authors.} 

\noindent {\small\bf Abstract} \quad  {\small {This paper provides a comprehensive review of the current methods and metrics used to evaluate the performance of Large Language Models (LLMs) in code generation tasks. With the rapid growth in demand for automated software development, LLMs have demonstrated significant potential in the field of code generation. The paper begins by reviewing the historical development of LLMs and their applications in code generation. Next, it details various methods and metrics for assessing the code generation capabilities of LLMs, including code correctness, efficiency, readability, and evaluation methods based on expert review and user experience. The paper also evaluates the widely used benchmark datasets, identifying their limitations and proposing directions for future improvements. Specifically, the paper analyzes the performance of code generation models across different tasks by combining multiple evaluation metrics, such as code compilation/interpretation success rates, unit test pass rates, and performance and efficiency metrics, to comprehensively assess the practical application of LLMs in code generation. Finally, the paper discusses the challenges faced in evaluating LLMs in code generation, particularly how to ensure the comprehensiveness and accuracy of evaluation methods and how to adapt to the evolving practices of software development. These analyses and discussions provide valuable insights for further optimizing and improving the application of LLMs in code generation tasks.}}

\vspace*{3mm}


\vspace*{4mm}

\end{CJK*}
\baselineskip=18pt plus.2pt minus.2pt
\parskip=0pt plus.2pt minus0.2pt
\begin{multicols}{2}

\section{Introduction}

Large Language Models (LLMs) have seen rapid development in recent years, garnering widespread attention in the academic community. By learning vast amounts of natural language data, LLMs are capable of producing fluent human language and have demonstrated the ability to understand semantics and perform complex tasks. With the continuous evolution of LLMs, they have exhibited outstanding capabilities in generating code, bringing new possibilities to automated software development~\cite{chen2021evaluating,bhattacharya2023exploring}.

The complexity of software systems continues to increase, and the demand for shorter development cycles is becoming increasingly urgent, making automated code generation an urgent need in the software industry. Traditional automatic code generation methods based on rules and pattern matching are no longer sufficient, while LLMs offer a new approach to this problem. LLMs can generate functional code based on natural language descriptions, greatly improving the efficiency of software development~\cite{fan2023large,min2023recent}.

Code generation tasks involve the automatic production of source code based on a given specification, which is often expressed in natural language or another high-level description. These tasks can include a wide range of activities, such as:
\begin{itemize}
    \item \textit{Translating natural language requirements into executable code}: For example, converting a user story or a feature description into a functional programming script.
    \item \textit{Filling in code templates}: Completing partially written code snippets by adding the missing parts based on the provided context.
    \item \textit{Refactoring and optimizing existing code}: Modifying and improving code structure and efficiency without altering its functionality.
    \item \textit{Generating test cases}: Creating test scripts and scenarios to validate the correctness and performance of a given piece of code.
\end{itemize}
By automating these tasks, LLMs can significantly streamline the software development process, reduce the likelihood of human error, and enable developers to focus on more complex and creative aspects of programming. As LLMs continue to be deeply integrated into the field of software engineering, it is necessary for us to fully understand their capabilities and limitations in code generation. This will not only help to improve the efficiency of software development but also promote further development of LLM technology in related fields~\cite{chen2021evaluating,ross2023programmer}.

However, comprehensively understanding the evaluation methods of LLMs in the field of code generation remains a challenge. This paper aims to provide a comprehensive and systematic review of the latest developments in the evaluation of LLMs for code generation, analyze the achievements and challenges in this area, and provide guidance for future research and applications. First, we will focus on discussing various methods and metrics for evaluating the performance of LLM code generation, including aspects such as code correctness, efficiency, and readability. In addition, we will summarize the widely used benchmark test suites in the industry and their limitations, and propose directions for future improvement. Finally, we will discuss the challenges faced by LLMs in the field of code generation, such as reliability, security, and interpretability, and look forward to future research prospects~\cite{liu2024no,rasnayaka2023empirical}.

The rest of this work is organized as follows: Section 2 focuses on the metrics for assessing the code generation capabilities of large language models. This section first discusses metrics such as code correctness, efficiency, and readability~\cite{liu2024no}. It then introduces evaluation methods based on expert review and user experience~\cite{cassano2023can}. These evaluation metrics and methods can comprehensively reflect the performance of large language models in code generation tasks. Section 3 reviews the current benchmarks and datasets used for evaluating code generation. This section begins with a summary of existing code generation benchmark suites, such as HumanEval, MBPP, and CodeXGLUE~\cite{ni2023l2ceval,yu2023codereval}. It then analyzes in detail the various metrics covered by these benchmark suites, including code correctness, efficiency, and readability~\cite{haque2022semantic}. The analysis indicates that while existing benchmarks and evaluation metrics have made certain progress, they still have some limitations, such as the difficulty in fully reflecting the performance of code in practical applications. Section 4 discusses the future challenges and opportunities in evaluating the code generation capabilities of large language models. This section first discusses the issue of scalability, that is, how to evaluate these models on a larger scale of data and more complex tasks~\cite{yan2023codescope}. Secondly, it analyzes the importance of multilingual generalization capabilities and how to evaluate the performance of large language models across a broader range of programming languages~\cite{cassano2022multipl}. Furthermore, security and robustness are also important aspects that must be considered in the evaluation of large language models~\cite{wang2023enhancing}. Finally, the section also explores how to evaluate the efficiency and usability of these models in practical applications~\cite{huang2024effibench}.

In summary, this paper provides a comprehensive reference for evaluating the performance of large language models in code generation tasks by systematically reviewing existing methods and metrics and looking forward to future development trends. Additionally, AutoSurvey~\cite{wang2024autosurvey} was utilized to retrieve relevant literature, ensuring a comprehensive overview. We hope that this review provides valuable insights for researchers and practitioners in the field, thereby fostering further advancements in this area.

\section{Code Generation Evaluation Metrics}

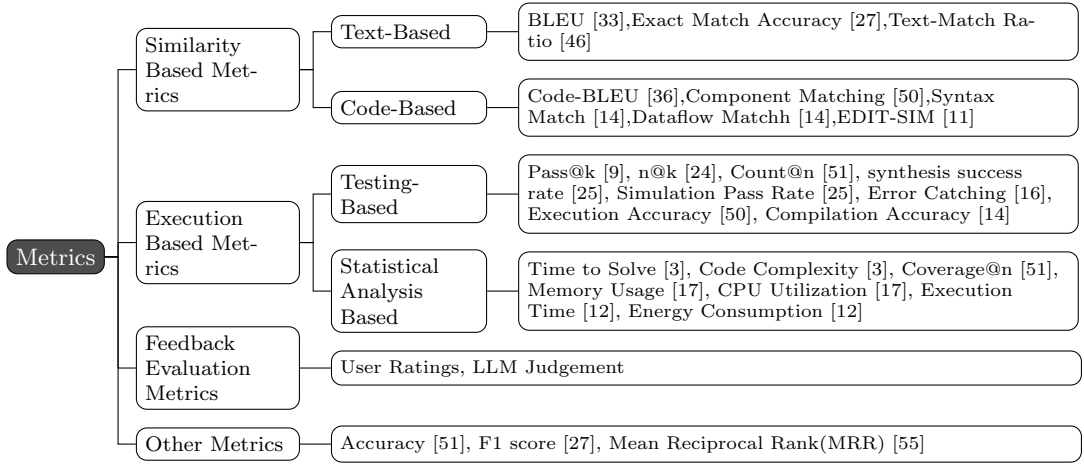
\begin{figure*}[t]
    \centering
    \begin{forest}
for tree={
forked edges,
draw,
rounded corners,
grow=east,
anchor=base west,
anchor=center,
reversed=true,
},
where level=0{font=\small}{},
where level=1{text width=5.5em,font=\footnotesize}{},
where level=2{text width=5.2em,font=\footnotesize}{},
where level=3{text width=1.8em,font=\scriptsize}{},
where level=4{text width=5.5em,font=\scriptsize}{},
[Metrics, fill=black, fill opacity=0.7, text=white   
    [Similarity Based Metrics, fill opacity=0.4
        [Text-Based, fill opacity=0.4
            [BLEU~\cite{papineni2002bleu}{,}Exact Match Accuracy~\cite{lu2021codexglue}{,}Text-Match Ratio~\cite{wu2024plot2code},
            text width=20.5em, fill opacity=0.4]
        ]
        [Code-Based,fill opacity=0.4
            [Code-BLEU~\cite{ren2020codebleu}{,}Component Matching~\cite{yu2019spider}{,}Syntax Match~\cite{haque2023fixeval}{,}Dataflow Matchh~\cite{haque2023fixeval}{,}EDIT-SIM~\cite{dibia2023alignjudge},
            text width=20.5em, fill opacity=0.4]
        ]
    ]
    [Execution Based Metrics, fill opacity=0.4
        [Testing-Based, fill opacity=0.4
            [Pass@k~\cite{chen2021humaneval}{,}
            n@k~\cite{Li_2022}{,}
            Count@n~\cite{zeng2024coderujb}{,}
            synthesis success rate~\cite{liu2023verilogeval}{,}
            Simulation Pass Rate~\cite{liu2023verilogeval}{,}
            Error Catching~\cite{hendrycksapps2021}{,}
            Execution Accuracy~\cite{yu2019spider}{,}
            Compilation Accuracy~\cite{haque2023fixeval},
            text width=20.5em, font=\scriptsize, fill opacity=0.4]
        ]
        [Statistical Analysis Based, fill opacity=0.4
            [Time to Solve~\cite{austin2021mbpp}{,}
            Code Complexity~\cite{austin2021mbpp}{,}
            Coverage@n~\cite{zeng2024coderujb}{,}
            Memory Usage~\cite{huang2024effibench}{,}
            CPU Utilization~\cite{huang2024effibench}{,}
            Execution Time~\cite{du2024mercury}{,}
            Energy Consumption~\cite{du2024mercury},
            text width=20.5em, fill opacity=0.4]
        ]
    ]
    [Feedback Evaluation Metrics,fill opacity=0.4
        [User Ratings{,}
        LLM Judgement,
        text width=27.7em, font=\scriptsize, fill opacity=0.4]
    ]
    [Other Metrics,  fill opacity=0.4
        [Accuracy~\cite{zeng2024coderujb}{,}
        F1 score~\cite{lu2021codexglue}{,}
        Mean Reciprocal Rank(MRR)~\cite{zhu2022xlcost},
        text width=27.7em, font=\scriptsize, fill opacity=0.4]
    ]
]
\end{forest}
    \caption{Classification of Evaluation Metrics for Code Generation Benchmarks.}
    \label{fig:Metrics}
\end{figure*}

\subsection{Evaluation Based on Similarity}

Similarity-based evaluation methods in code generation primarily assess the quality of generated code by comparing its similarity to reference code. Figure \ref{fig:Metrics} provides a classification of various evaluation metrics used in code generation benchmarks. These similarity evaluation methods are generally divided into following categories: similarity based metrics, execution based metrics and feedback based metrics. I will discuss several representative metrics in detail in the following sections.

\subsubsection{Traditional Similarity Metrics}

Traditional similarity metrics, initially used in the field of natural language processing, have also been applied to the evaluation of code generation capabilities. These methods assess the quality of generated code by calculating its similarity to reference code. Common similarity metrics include BLEU, ROUGE, and METEOR, with the CodeXGLUE dataset utilizing the BLEU metric to calculate the similarity between generated and correct code.

\textbf{BLEU (Bilingual Evaluation Understudy)} is a metric used for machine translation evaluation, first proposed by Papineni et al~\cite{papineni2002bleu}. It measures the overlap of n-grams between the generated text and reference text. Specifically, BLEU calculates the number of matching n-grams between the generated text and reference text, then averages these matches with weights to produce an overall similarity score.

In the context of code generation, BLEU is used to evaluate the similarity between generated and reference code. Although BLEU excels in natural language processing, it also has some issues when applied to code generation. Firstly, the syntax and semantic structure of code are more complex than natural language, so relying solely on n-gram matching may not accurately measure code similarity. Secondly, code often contains a large number of unique identifiers such as variable names and function names, which may be completely different across different code snippets but actually perform the same function.

Nevertheless, due to its simplicity and intuitiveness, BLEU is still widely used for preliminary evaluation in code generation. For example, the CodeXGLUE dataset~\cite{lu2021codexglue} uses BLEU as one of its main evaluation metrics to measure the similarity between generated and correct code. This indicates that, despite its flaws, BLEU remains a useful tool, especially when more complex alternatives are not available.

Besides BLEU, \textbf{ROUGE (Recall-Oriented Understudy for Gisting Evaluation)} and \textbf{METEOR (Metric for Evaluation of Translation with Explicit ORdering)} are two traditional similarity metrics initially used in natural language processing that have been adapted for code generation evaluation. ROUGE focuses on recall, measuring the frequency of n-grams from the reference text appearing in the generated text, which can capture subtle differences when the generated code covers the logical steps of the reference code despite different variable and function names. METEOR, on the other hand, combines precision and recall while considering lexical matching, multi-word matching, and semantic relationships. Its nuanced approach to lexical matching, including synonyms and morphological variations, makes it more flexible for code evaluation, accurately reflecting the quality of generated code even when different variable and function names are used. Both metrics complement each other and help address some of the limitations of BLEU in evaluating the similarity and quality of generated code.

\textbf{Exact Match (EM)}~\cite{lu2021codexglue} metric is a straightforward and stringent evaluation method used to measure the accuracy of generated code by directly comparing it to the reference code. This metric is particularly useful in assessing the overall correctness of the code completion process, taking into account elements such as identifiers, keywords, operators, delimiters, and literals.
Exact Match (EM) calculates the percentage of generated code snippets that exactly match the reference code snippets. This means that for a generated code snippet to be considered a match, it must be identical to the reference code in every aspect, including syntax, structure, and content. It provides a clear and unambiguous measure of code generation accuracy, making it a valuable tool for evaluating code completion models.

\textbf{Edit Distance}~\cite{editdistance}, also known as Levenshtein Distance, is a widely used metric for evaluating the similarity between two sequences by measuring the minimum number of single-character edits required to transform one sequence into the other. These edits include insertions, deletions, and substitutions. In the context of code generation, Edit Distance can be employed to assess how closely the generated code matches the reference code by quantifying the effort needed to convert one into the other. Edit Distance is calculated using the following recursive definition:

{\scriptsize
\begin{align}
\text{lev}(a, b) = 
\begin{cases} 
|a| & \text{if } |b| = 0 \\
|b| & \text{if } |a| = 0 \\
\text{lev}(\text{tail}(a), \text{tail}(b)) & \text{if } \text{head}(a) = \text{head}(b) \\
1 + \min \begin{cases} 
\text{lev}(\text{tail}(a), b) \\
\text{lev}(a, \text{tail}(b)) \\
\text{lev}(\text{tail}(a), \text{tail}(b)) 
\end{cases} & \text{otherwise}
\end{cases}
\end{align}}

The \(\text{tail}(x)\) denotes the string obtained by removing the first character of \(x\) (i.e., \(\text{tail}(x_0x_1 \ldots x_n) = x_1x_2 \ldots x_n\)), and \(\text{head}(x)\) represents the first character of \(x\) (i.e., \(\text{head}(x_0x_1 \ldots x_n) = x_0\)). In the minimum operation, the first term corresponds to deletion (from \(a\) to \(b\)), the second to insertion, and the third to replacement.
The Edit Distance metric is particularly useful for code generation evaluation as it provides a more flexible measure of similarity compared to Exact Match. It accounts for minor differences and variations in the code, such as different variable names or slight changes in syntax, while still reflecting the overall effort needed to achieve an exact match. This makes it a valuable complement to other metrics like BLEU, ROUGE, METEOR, and Exact Match, offering a more nuanced understanding of the generated code's quality.

\subsubsection{Code-Specific Similarity Metrics}

To more accurately assess the quality of code generation, researchers have developed similarity metrics specifically tailored for code. These methods introduce programming-specific characteristics, such as Abstract Syntax Trees (ASTs), data flow, and token matching, to provide a more comprehensive evaluation of the syntactic and semantic similarity between generated and reference code.

\textbf{CodeBLEU} is an evaluation method specifically for code that extends the traditional BLEU metric~\cite{ren2020codebleu}. CodeBLEU combines n-gram matching from BLEU while introducing syntactic and semantic information of code. Specifically, it includes weighted n-gram matching, syntactic AST matching, and semantic data flow matching. Each component calculates a score, which is then combined in a weighted manner to produce the total score. The weighted n-gram matching is an extension of the traditional BLEU algorithm, assigning different weights to different n-grams to better reflect the keywords and structures in code. Syntactic AST matching uses Abstract Syntax Trees (ASTs) to compare the syntactic structure of candidate and reference code, calculating scores by matching subtrees. Semantic data flow matching evaluates the semantic similarity of code by analyzing the data flow diagrams within the code. Such a design enables CodeBLEU to capture not only the surface similarity of code but also to deeply understand the internal logic and functionality, improving the accuracy and comprehensiveness of the evaluation. In summary, when calculating similarity, CodeBLEU analyzes the AST structures of both the generated and reference code, ensuring that not only the textual content is similar but also that the code structure and logic are as close as possible. This allows CodeBLEU to more fully consider the syntactic and semantic similarity of code when evaluating the quality of code generation, thereby providing a more accurate assessment.

CodeBLEU was used to evaluate the task of generating code from natural language descriptions, and the results showed that the Pearson correlation coefficient between CodeBLEU and programmer scoring was higher than that of BLEU and accuracy metrics~\cite{ren2020codebleu}. The application of CodeBLEU is not limited to a single programming language; it performs excellently across multiple programming languages~\cite{ren2020codebleu}.

\textbf{Other Similarity Methods} include metrics based on data flow analysis and semantic similarity metrics. Data flow analysis assesses code quality by comparing the similarity of data flows between generated and reference code, providing a deeper understanding of the code's semantics. Data-aware techniques analyze variables and data flows in generated code to verify functional correctness, ensuring the generated code achieves expected functionality~\cite{cotroneo2024automating}. These metrics are also used to evaluate code optimization and repair, significantly enhancing code performance~\cite{yang2023asteria}.

Semantic similarity metrics focus on the actual functionality and behavior of the code. For instance, in code summarization tasks, semantic similarity metrics evaluate the quality of the summary by measuring the semantic similarity between the generated code summary and the reference summary~\cite{haque2022semantic}. Another example is DeepSemantic, which utilizes deep learning models to generate semantic representations of binary code for code similarity measurement, showing potential in cross-architecture vulnerability detection and patch analysis~\cite{koo2106semantic}.

\subsection{Execution-Based Evaluation}

\subsubsection{Compilation/Interpretation Success Rate}
The compilation or interpretation success rate is a crucial metric for evaluating the quality of code generation, assessing whether the generated code can be successfully compiled or interpreted without syntactic errors~\cite{siddiq2023lightweight,wang2022compilable}. A high compilation or interpretation success rate indicates that the code adheres to the syntactic rules of the programming language, which is a fundamental requirement for any functional code. If the code cannot be successfully compiled or interpreted, it cannot be executed further and thus cannot achieve its intended functionality.

To evaluate the compilation or interpretation success rate, we typically use standard compilers and interpreters for various programming languages, such as GCC for C/C++, the Python interpreter, etc. These tools can verify the syntactic correctness of the generated code and prepare it for execution. Through these tools, we can directly assess the compilation or interpretation success rate of the generated code, thereby gaining a basic understanding of the performance of the code generation model. For instance, the FRANC framework significantly improves the proportion of generated code that passes compilation through the use of static filters, enhancing the quality of Java suggestions by 9\% to 46\% and Python suggestions by 10\% to 43\%~\cite{siddiq2023lightweight}. COMPCODER proposes a three-stage pipeline that uses compiler feedback to generate compilable code~\cite{wang2022compilable}. Its pipeline includes language model fine-tuning, compilability enhancement, and compilability discrimination. This method not only improves the successful compilation rate but also makes the generated code more reliable in practical applications.

\subsubsection{Unit Test Pass Rate}

Unit testing is an important metric for evaluating code quality, which verifies the correctness of the code by running the generated code with predefined test cases~\cite{zhong2023codegen,ni2023lever,ren2023you}. This method is crucial for assessing the expected performance of the code under various conditions, as it ensures the practical utility and reliability of the code. For instance, Humaneval is a representative unit testing framework, and its Pass@k metric has become a classic evaluation metric for the code generation capabilities of large language models (LLMs)~\cite{chen2021humaneval}. The Pass@k metric measures the probability of the generated code passing the test within the first k attempts, effectively evaluating the performance and reliability of the code generation model. 

By systematically integrating unit testing steps and using error feedback to iteratively correct the generated code, the unit test pass rate of the generated code can be significantly improved, ensuring the reliability and stability of the code in practical applications. For example, CodeGen-Test adds a program testing step during the code generation process, combining testing information to iteratively produce code that meets functional requirements~\cite{zhong2023codegen}. LEVER utilizes execution results to detect and correct erroneous programs, continuously improving the quality of the generated code~\cite{ni2023lever}. Furthermore, the Multi-Stage Generation Process introduced by VCP transforms verification errors into specific hints, guiding the model to regenerate outputs that address the discovered errors. This process significantly reduces the error rate and improves generation quality~\cite{ren2023you}.

\subsubsection{Performance and Efficiency Evaluation}
Performance and efficiency evaluation refers to the assessment of the actual runtime performance of generated code by measuring its time and space complexity. Efficient code is crucial for practical applications, and performance evaluation helps identify potential bottlenecks and optimize the code. In software development, optimizing computational efficiency, in addition to ensuring functional correctness, is a universal and significant objective. Efficient code enhances system performance and plays a more substantial role in resource-constrained environments. Therefore, focusing on improving the efficiency and performance of code during the development and evaluation of code generation models is essential.

By integrating performance and efficiency evaluation into the code generation and testing process, we can ensure that the generated code is not only functionally correct but also performs well in practical applications, providing foundational data support for optimization. EffiBench~\cite{huang2024effibench} and Mercury~\cite{du2024mercury} are notable frameworks in this domain. EffiBench is a benchmarking framework for evaluating the efficiency of automatically generated code, encompassing a variety of programming tasks and languages. Mercury, on the other hand, is a specialized benchmarking framework designed to assess the efficiency of code generated by Large Language Models (LLMs).

\subsection{Feedback-Based Evaluation}

Feedback-based evaluation methods are essential for comprehensively assessing the quality of generated code, as they incorporate human judgment and expertise to evaluate various aspects of code quality. These methods often involve blind peer review, real-world application evaluation, readability evaluation, and maintainability evaluation.

\subsubsection{Blind Peer Review}
Blind peer review is a common and effective method for evaluating code quality comprehensively. In this method, reviewers assess code snippets generated by different models without knowing the identity of the models, selecting the superior code based on predetermined criteria. This approach eliminates potential biases, making the evaluation results more objective and fair. For instance, in the MT-Bench study~\cite{zheng2024judging}, reviewers conducted multiple rounds of comparative evaluations based on criteria such as functionality, clarity, and maintainability. Multiple reviewers and rounds of comparison ensure fairness and consistency, providing detailed insights into the strengths and weaknesses of different code generation models.

\subsubsection{Real-World Application Evaluation}
Another important evaluation method is to deploy the generated code in actual application environments and assess its performance in real-world tasks. This method fully evaluates the practicality and reliability of the code, reflecting its real-world effectiveness. Generated code is applied to real programming tasks, with metrics such as error rate, debugging time, and maintenance cost recorded. This approach provides valuable feedback on the code's functionality, stability, and adaptability. For example, generated code might perform excellently in a controlled environment but face performance bottlenecks or compatibility issues in practical applications. Real-world application evaluation helps identify and address these issues, thereby improving the overall quality of the generated code~\cite{ccano2020human}.

\subsubsection{Readability Evaluation}
The readability of code is crucial for understanding and maintaining it. Human evaluation methods focus on assessing the functionality, clarity, and maintainability of the code. Reviewers consider naming conventions, comments, and code logic to determine clarity and conciseness. Clear and concise code improves development efficiency and long-term sustainability. For example, reviewers check if variable and function names are descriptive, if appropriate comments explain the code logic, and if the code structure is easy to understand~\cite{wang2023leti}.

\subsubsection{Maintainability Evaluation}
Maintainability refers to the ease with which code can be updated and modified in the future. Code with high maintainability should have good modular design~\cite{lahiri2022interactive}, detailed documentation~\cite{wout2019learning}, and adherence to programming standards. Modular design makes code easier to modify and extend by dividing it into independent, reusable modules, each responsible for a specific function~\cite{lahiri2022interactive}. Reviewers evaluate whether the code is reasonably divided into such modules. Additionally, they check for comprehensive documentation and comments, such as descriptions of functions and classes, parameters, and return values. For instance, each function should have detailed comments explaining its functionality, input parameters, and return values. Good documentation and comments help current and future developers understand and maintain the code~\cite{wout2019learning}.

\section{Code Generation Evaluation Benchmarks}

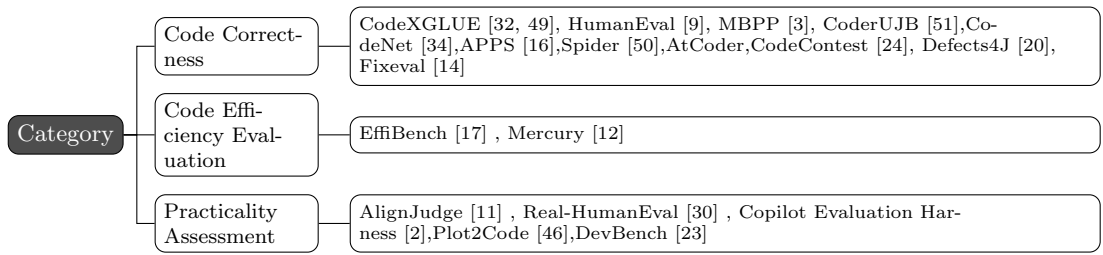
\begin{figure*}[t]
    \centering
    \begin{forest}
for tree={
forked edges,
draw,
rounded corners,
grow=east,
anchor=base west,
anchor=center,
reversed=true,
},
where level=0{font=\small}{},
where level=1{text width=5.5em,font=\footnotesize}{},
where level=2{text width=5.2em,font=\footnotesize}{},
where level=3{text width=1.8em,font=\scriptsize}{},
where level=4{text width=5.5em,font=\scriptsize}{},
[Category, fill=black, fill opacity=0.7, text=white   
    [Code Correctness,  fill opacity=0.4
        [CodeXGLUE~\cite{ni2023l2ceval,yu2023codereval}{,} HumanEval~\cite{chen2021humaneval}{,} MBPP~\cite{austin2021mbpp}{,} CoderUJB~\cite{zeng2024coderujb}{,}CodeNet~\cite{puri2021codenet}{,}APPS~\cite{hendrycksapps2021}{,}Spider~\cite{yu2019spider}{,}AtCoder{,}CodeContest~\cite{Li_2022}{,}
        Defects4J~\cite{defects4J2014}{,}
        Fixeval~\cite{haque2023fixeval}
,
        text width=27.7em,font=\scriptsize,  fill opacity=0.4]
    ]
    [Code Efficiency Evaluation,  fill opacity=0.4
        [EffiBench~\cite{huang2024effibench}
{,} Mercury~\cite{du2024mercury},
        text width=27.7em,font=\scriptsize, fill opacity=0.4]
    ]
    [Practicality Assessment,  fill opacity=0.4
        [AlignJudge~\cite{dibia2023alignjudge}
{,} Real-HumanEval~\cite{mozannar2024realhumaneval}
{,} Copilot Evaluation Harness~\cite{agarwal2024copilot}{,}Plot2Code~\cite{wu2024plot2code}{,}DevBench~\cite{li2024devbench},
        text width=27.7em,font=\scriptsize, fill opacity=0.4]
    ]
]
\end{forest}
    \caption{Classification of Code Generation Benchmarks.}
    \label{fig:Category}
\end{figure*}

Evaluating the performance of code generation models is a multifaceted task that involves various benchmarks designed to test different aspects of code quality. Figure \ref{fig:Category} presents a classification of these benchmarks, categorizing them based on the specific evaluation criteria they address. These benchmarks are essential for understanding how well a model performs in generating accurate, efficient, and practical code. I will delve into several representative benchmarks in the following sections.

% In the following sections, I will delve into several representative benchmarks, discussing their dataset composition, testing methods, performance metrics, and analysis of test results. This detailed exploration aims to provide a comprehensive understanding of the strengths and limitations of current code generation models.

\subsection{Code Correctness}
In the field of code generation, testing for code correctness is an essential task. It not only helps evaluate the quality of the code generated by models but also provides feedback for model optimization. In this chapter, we will introduce five datasets used for code correctness testing: CodeXGLUE~\cite{lu2021codexglue}, HumanEval~\cite{chen2021humaneval}, MBPP~\cite{austin2021mbpp}, CoderUJB~\cite{zeng2024coderujb}, and VerilogEval~\cite{liu2023verilogeval}. These methods each have their own characteristics in terms of dataset composition, testing methods, performance metrics (such as pass@k), and analysis of test results.

\subsubsection{CodeXGLUE}
CodeXGLUE encompasses various code understanding and generation tasks using multiple large datasets. For example, code clone detection utilizes BigCloneBench and POJ-104~\cite{clf2014-1} datasets, defect detection uses the Devign~\cite{2019Devign} dataset, text-to-code generation employs the CONCODE~\cite{2018CONCODE} dataset, and code summary generation relies on the CodeSearchNet~\cite{2019CodeSearchNet} dataset. These datasets span multiple programming languages, including Java, C/C++, and Python, ensuring broad applicability and representativeness of CodeXGLUE's evaluation results.

The datasets in CodeXGLUE have undergone rigorous preprocessing and filtering to ensure high data quality and consistency, supporting reliable and reproducible evaluation results. Their diversity and extensive coverage enhance the model's adaptability in practical applications, making CodeXGLUE valuable in both academic research and industry.

Performance metrics used in CodeXGLUE include BLEU, Exact Match Accuracy, F1 score, and CodeBLEU. These metrics cover traditional evaluation methods and introduce code-specific standards to better reflect the quality of code generation and understanding. For instance, Exact Match and CodeBLEU are used in text-to-code generation, while Accuracy and F1 score are used in code clone detection.

Benchmark evaluations demonstrate the effectiveness of pre-trained models in CodeXGLUE. CodeGPT achieves a CodeBLEU score of 35.98 in code generation, indicating strong generation capabilities. CodeBERT performs well in code clone detection and defect detection, with high Accuracy and F1 scores. These results highlight the advantages of pre-trained models in code tasks and their potential for further improvement through multi-task and transfer learning.

Overall, CodeXGLUE provides a comprehensive evaluation framework for code understanding and generation tasks, helping researchers identify and optimize suitable models for specific tasks. It promotes technological progress and innovation in the field, offering valuable references for future research and model improvement. With ongoing research and the introduction of more datasets, CodeXGLUE is expected to support the continuous development and enhancement of code generation technology.

\subsubsection{HumanEval}
\begin{figure}[H]
    \includegraphics[width=\columnwidth]{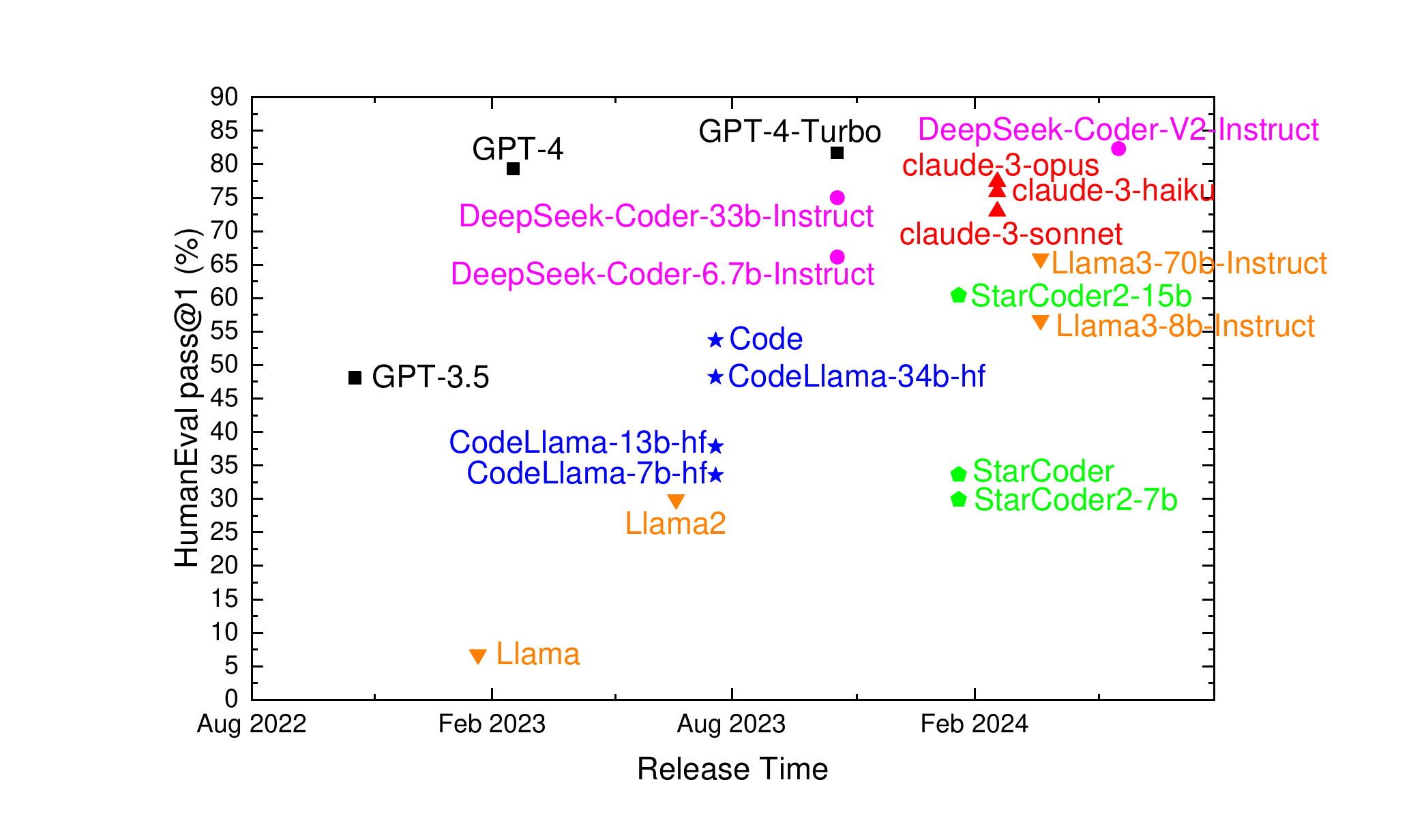}
    \caption{Pass@1 Performance of LLMs on HumanEval Over Time.}
    \label{fig:HumanEval}
\end{figure}

The HumanEval dataset is used to assess the practical performance of code generation models. It comprises 164 Python programming tasks, each with a natural language description and corresponding test cases. Multiple test cases are designed for each task to ensure comprehensive functionality coverage and correctness of the generated code. These tasks span various programming concepts, from basic control structures to complex algorithms and data structures, thoroughly testing the capabilities of code generation models. The hand-written nature of these tasks ensures quality and uniqueness, avoiding issues from programmatically copied tasks.

The main performance metric of HumanEval is the pass rate (Pass@k), the proportion of at least one of the top k generated code snippets passing all test cases. Pass@1, Pass@5, and Pass@10 are commonly used metrics. By comparing pass rates, the relative advantages and disadvantages of different models can be assessed. Pass@1 reflects the model's ability to generate high-quality code on the first attempt, while Pass@5 and Pass@10 reflect performance in diversity.

Analysis of HumanEval test results shows that GPT based models perform well, with higher Pass@1 scores than traditional methods. This indicates effective performance of pretrained GPT in real programming tasks, particularly in code correctness and functionality. Detailed analysis of different models' performances on various tasks helps identify strengths and weaknesses, guiding further optimization. For instance, CodeX excels in string processing tasks but may require optimization for complex algorithms. In-depth analysis also reveals differences in handling various natural language descriptions, offering insights for improving description expressions. 

Figure \ref{fig:HumanEval} illustrates the Pass@1 performance of representative LLMs on the HumanEval dataset over time, highlighting the improvements in code generation capabilities.

In summary, the HumanEval dataset and testing method provide important references for evaluating and improving code generation models, guiding future research. Comparing Pass@k scores identifies strengths and weaknesses, facilitating targeted improvements.

\subsubsection{MBPP (Mostly Basic Python Problems)}

\begin{figure}[H]
    \includegraphics[width=\columnwidth]{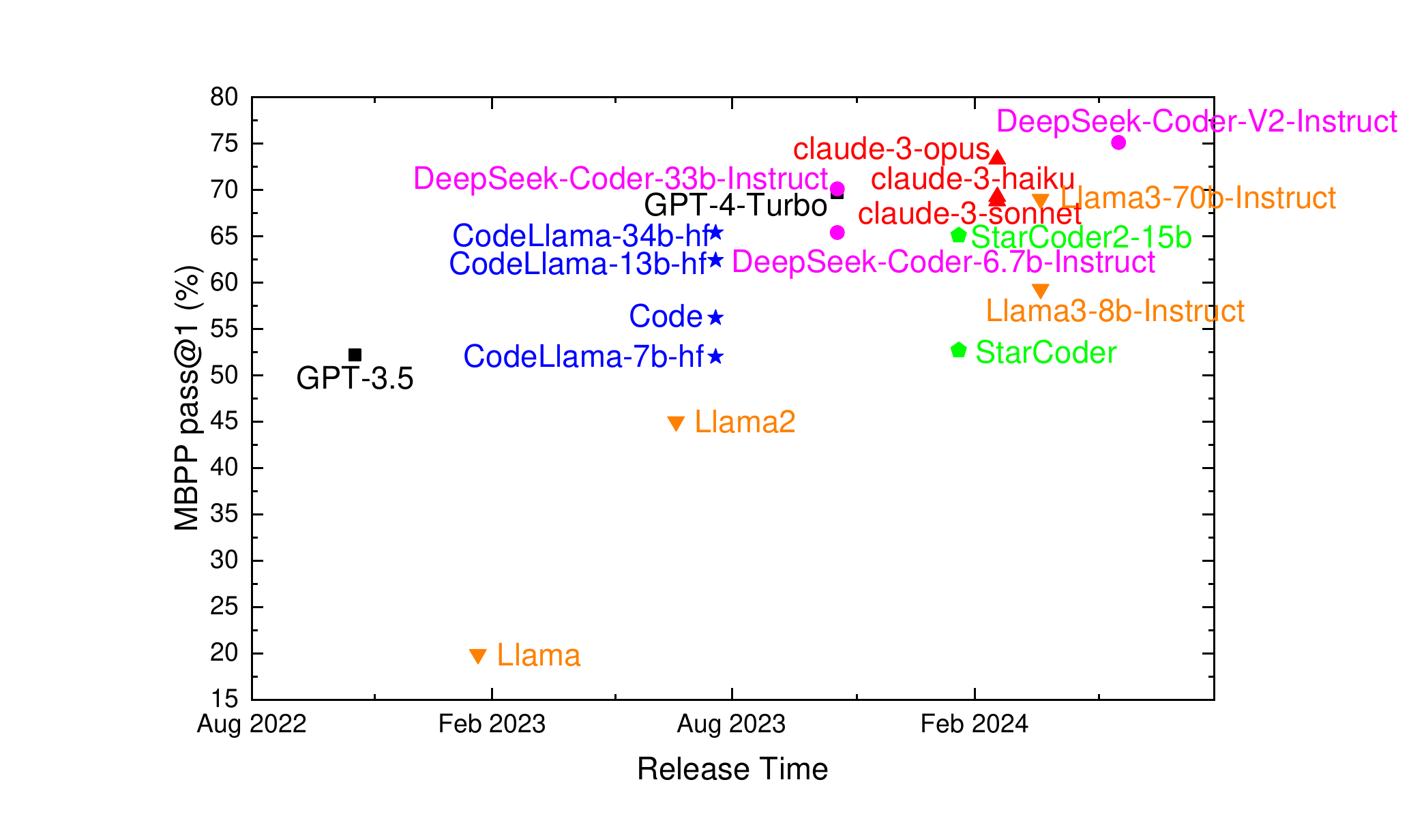}
    \caption{Pass@1 Performance of LLMs on MBPP Over Time.}
    \label{fig:MBPP}
\end{figure}

The MBPP dataset consists of 500 Python programming problems, each with a natural language description, example input-output pairs, and solution code, divided into training, validation, and test sets. It covers a wide range of programming concepts, from basic string operations to complex algorithms and data structures. The dataset is designed to test the model's performance in practical programming environments. Each problem is crafted and reviewed to ensure quality and representativeness, with clear and concise descriptions to aid in task understanding.

Tasks in MBPP test the model's ability to generate correct code and understand natural language descriptions. The training set provides some learning resource aids models in learning to generate high-quality code, while the validation and test sets assess the model's generalization ability and performance on new problems. Comprehensive test cases cover functionality, boundary conditions, and exception handling, ensuring thorough evaluation.

The MBPP testing method requires the generated code to pass all provided test cases, evaluating correctness and robustness under various scenarios. Performance metrics such as time and resource consumption are also recorded to evaluate efficiency. The primary performance metric is Pass@k, indicating the proportion of correct code generated within the first k attempts, with Pass@1, Pass@5, and Pass@10 being commonly used. Additionally, metrics like average time to solve problems and the complexity of generated code assess practicality, efficiency, and code quality.

Analysis of MBPP test results shows that Transformer-based models perform exceptionally well, particularly in Pass@1 and Pass@5 metrics, indicating high efficiency and accuracy. Detailed analysis helps identify strengths and weaknesses, guiding further optimization. For instance, Transformer models excel in string operations but may require optimization for complex algorithms. Differences in handling various natural language descriptions also offer insights for improvement. 

Figure \ref{fig:MBPP} illustrates the Pass@1 performance of representative LLMs on the MBPP dataset over time, showcasing the remarkable improvements in the capabilities of code generation models.

In summary, MBPP's comprehensive testing methods and detailed result analysis enhance understanding and evaluation of code generation models, supporting future research and applications.

\subsubsection{CoderUJB}

CoderUJB is a comprehensive Java benchmark test set designed to assess the performance of Large Language Models (LLMs) in various programming tasks and real software development scenarios. Unlike HumanEval and MBPP, which focus on Python, CoderUJB includes 2,239 programming problems extracted from 17 open-source Java projects. These cover five tasks: 238 functional code generation problems, 140 code-based test generation problems, 451 issue-based test generation problems, 470 automatic program repair problems, and 940 defect detection problems, each with complete project context.

CoderUJB's testing method involves multiple steps: task allocation, code generation, unit testing, and comprehensive assessment. Task allocation assigns the model specific programming tasks, simulating real development scenarios. During code generation, the model generates the function body based on the provided function signature and comments. The generated code must pass a compilation check to ensure syntactic correctness.

Unit testing verifies the quality of code generation. Preset test cases cover various inputs and boundary conditions, ensuring the generated code's correctness and robustness. The pass rate of these tests is the main indicator of code quality. Additionally, multi-task performance, execution efficiency, and code quality are evaluated to ensure the model's applicability in diverse development scenarios.

CoderUJB uses refined evaluation metrics: Pass@k, count@n, coverage@n, and accuracy. Pass@k measures the probability of generating correct code in k attempts, while count@n quantifies the number of successful generations within n attempts. Coverage@n assesses how many test cases the generated code covers, and accuracy measures the proportion of code that passes all test cases. These metrics provide a comprehensive assessment framework for comparing model performance in programming tasks.

Results show that while LLMs perform well in code generation tasks, challenges remain in non-functional tasks like test generation and defect detection. Continuous pre-training and instruction fine-tuning have mixed effects, indicating the need for careful strategy selection. Comprehensive assessments highlight the varying performance of models across different tasks, emphasizing the need for meticulous strategies to enhance LLM capabilities in software engineering.

In summary, CoderUJB provides a realistic and comprehensive framework for programming capability assessment in Java, offering valuable insights for the future development of LLMs in software engineering. This research demonstrates the potential of current LLMs while identifying key challenges for practical applications, guiding future improvements in model training and assessment methods.

\subsubsection{VerilogEval}

VerilogEval is a dataset dedicated to the code generation and verification of Verilog, a hardware description language. Unlike HumanEval and MBPP, which focus on software, VerilogEval targets hardware design tasks, ensuring the generated code's effectiveness in synthesis and simulation. The dataset includes tasks covering combinational logic circuits, sequential logic circuits, and state machine design, with each task providing detailed natural language descriptions and design constraints like timing, power, and area.

The testing method involves synthesis and simulation. The model-generated Verilog code is first checked for syntactic correctness through synthesis tools and then tested for functionality via simulation. This method ensures that the generated code is not only syntactically correct but also functionally robust. The synthesis time and simulation time are recorded to evaluate efficiency, which is crucial in hardware design.

Performance metrics for VerilogEval include synthesis success rate, simulation pass rate, and design performance (timing, power, and area). High synthesis and simulation pass rates indicate basic correctness, while excellent design performance reflects high efficiency and resource utilization. These metrics provide a comprehensive assessment of the model's capabilities in practical hardware design applications.

Analysis of VerilogEval results shows that deep learning-based code generation models perform well in synthesis and simulation but need improvement in design performance. Simple tasks like combinational logic circuits show high success rates, whereas complex tasks like sequential logic circuits and state machines present challenges. Detailed analysis helps identify strengths and weaknesses, guiding further optimization.

Overall, VerilogEval demonstrates the strong potential of deep learning models in hardware design, providing a solid foundation for future research. The dataset's comprehensive testing methods and detailed result analysis offer valuable insights for model evaluation and improvement, supporting the development of more reliable and efficient code generation models for hardware design.

\subsubsection{Other Benchmarks}

In addition to the methods previously detailed, there are numerous other innovative approaches and metrics for evaluating code generation correctness in large language models.

\begin{itemize}
    \item \textbf{CodeNet}: A large-scale dataset developed by IBM, containing over 14 million code samples and around 5000 problems in 55 programming languages. It is designed to support tasks such as code classification, code completion, and code similarity analysis.
% Project CodeNet: A Large-Scale AI for Code Dataset for Learning a Diversity of Coding Tasks

\item \textbf{APPS (Automated Programming Progress Standard)}: This benchmark contains a variety of coding problems designed to measure the problem-solving capabilities of AI models. It includes simple to complex problems with detailed performance metrics.
% Evaluating Large Language Models Trained on Code
\item \textbf{Spider}: A complex and cross-domain benchmark for evaluating text-to-SQL models. It includes thousands of natural language questions and corresponding SQL queries across multiple domains, ensuring a robust assessment of model performance in generating correct SQL code.
% Spider: A Large-Scale Human-Labeled Dataset for Complex and Cross-Domain Semantic Parsing and Text-to-SQL Task
% \item \textbf{JuPyTEx (Jupyter Notebook Text Execution)}: This benchmark focuses on evaluating the ability of models to generate and execute code within Jupyter Notebooks. It includes tasks like cell completion, error correction, and context-aware code generation.
% JuPyTEx: A Benchmark for Executable Document Understanding and Generation

\item \textbf{AtCoder}: This dataset comprises coding problems and their solutions from the AtCoder programming contest platform. It includes a wide range of problem difficulties and multiple programming languages, providing a thorough evaluation of code generation models.
% Programming Competitions for Assessing AI Capabilities
\item \textbf{CodeContest}: This benchmark includes problems from various competitive programming platforms like Codeforces, LeetCode, and HackerRank. It is used to evaluate the performance of models in generating correct and efficient code solutions under competitive constraints.
% Evaluating AI Systems for Code Generation in Competitive Programming

\item \textbf{Defects4J}: A benchmark for evaluating models on real-world Java programs. It contains over 800+ bug-fix pairs from popular open-source projects, providing a rigorous test for bug detection and repair capabilities.

\item \textbf{Fixeval}: A benchmark comprising of buggy code submissions to competitive programming problems and their corresponding fixes. 

\end{itemize}

% Defects4J: A Database of Existing Faults to Enable Controlled Testing Studies for Java Programs
These benchmarks cover a wide range of programming languages and problem types, providing a comprehensive assessment framework for evaluating code generation models in various real-world scenarios.

\subsection{Code Efficiency Evaluation}

In the realm of code generation, the efficiency assessment of code is a critical aspect, directly influencing the feasibility and value of the generated code in practical applications. Efficient code not only conserves computational resources but also enhances user experience and system response speed. Therefore, evaluating the efficiency of generated code is an essential step in ensuring code quality. This chapter will introduce three methods for assessing code efficiency: EffiBench~\cite{huang2024effibench}, Mercury~\cite{du2024mercury}. Through these methods, we can gain a more comprehensive understanding of the efficiency of generated code, promoting the development of more efficient code generation models.

\subsubsection{EffiBench}

\begin{figure}[H]
    \includegraphics[width=\columnwidth]{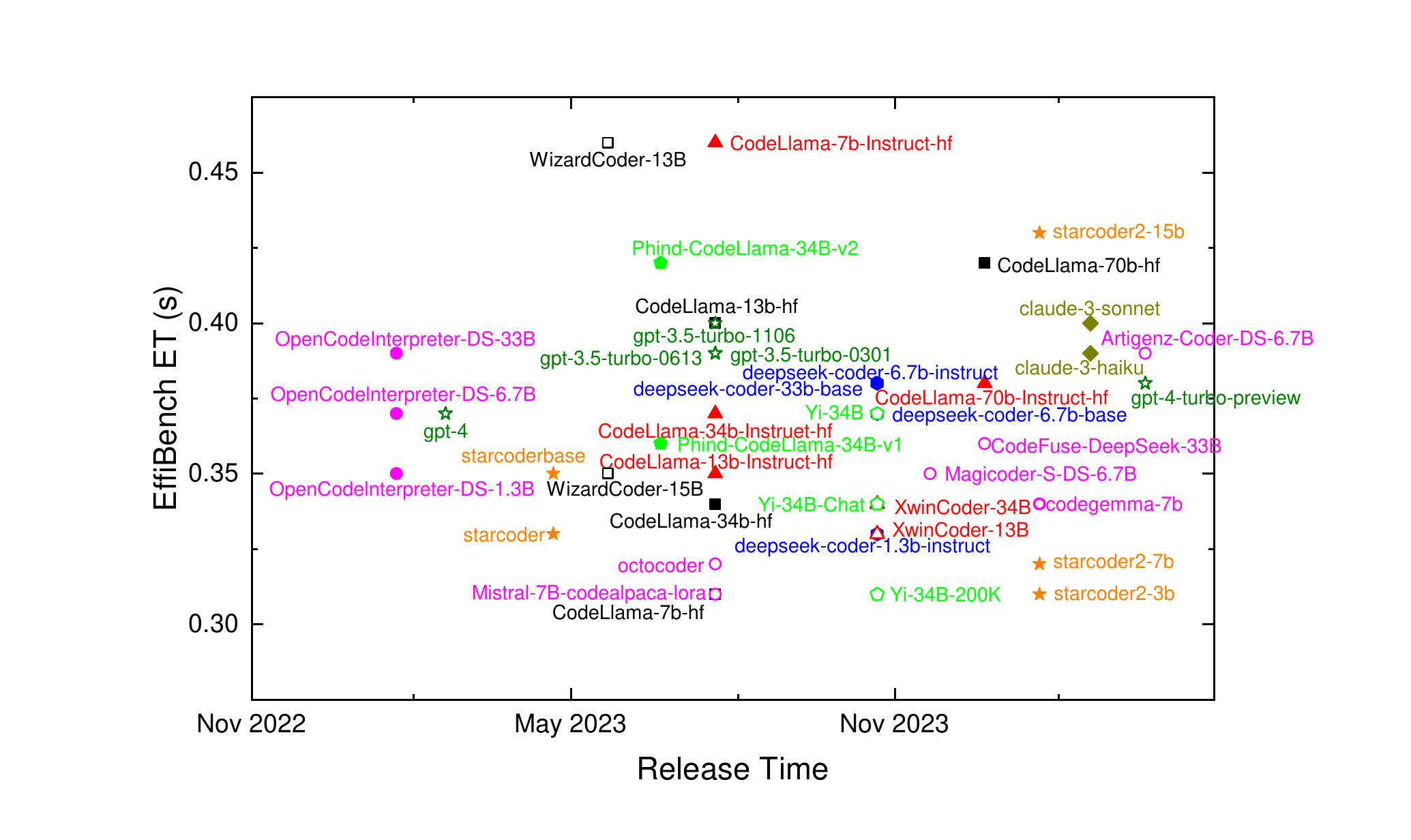}
    \caption{Execution Time (ET) Evaluation of LLMs on EffiBench Over Release Time.}
    \label{fig:EffiBenchet}
\end{figure}

\begin{figure}[H]
    \includegraphics[width=\columnwidth]{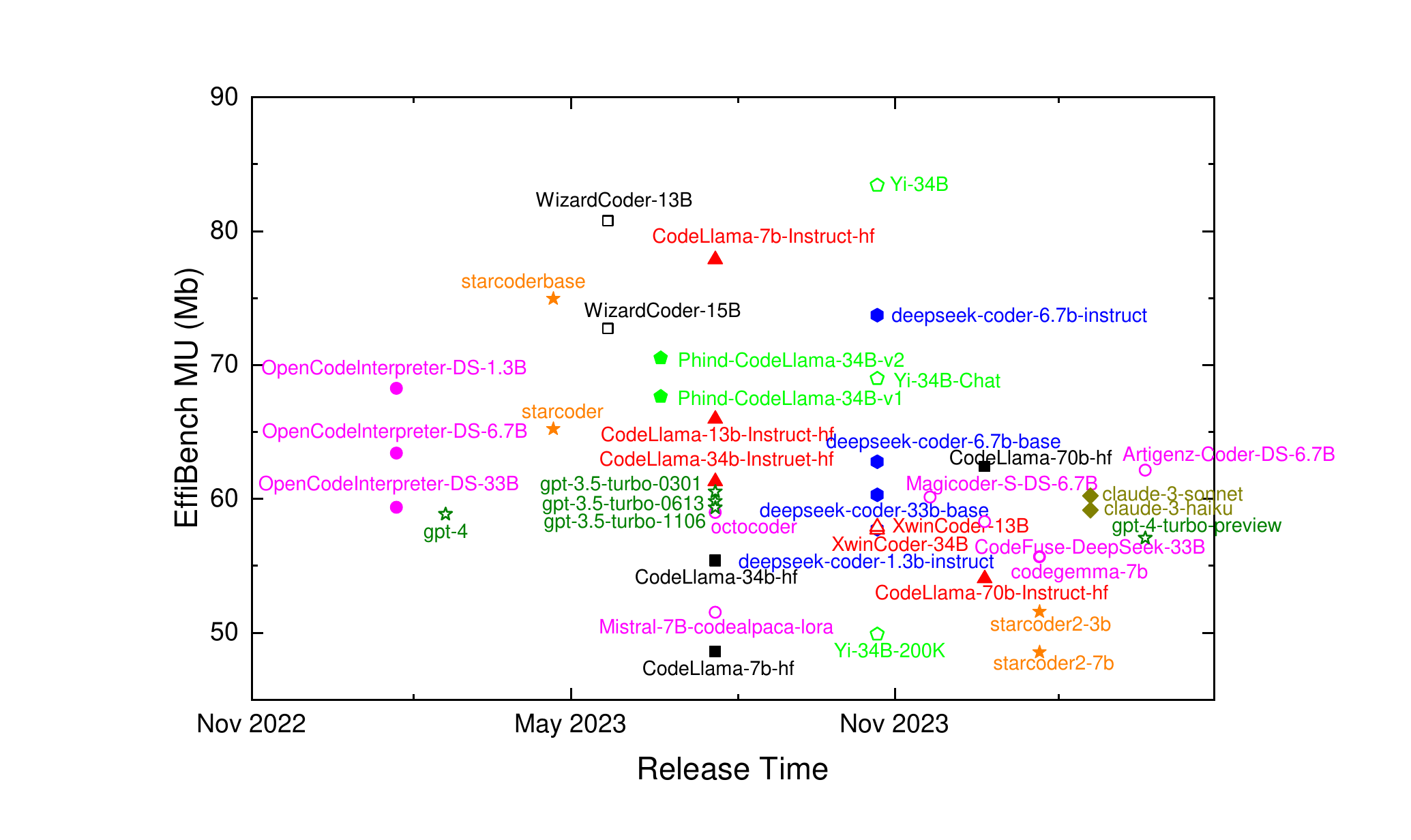}
    \caption{Memory Usage (MU) Evaluation of LLMs on EffiBench Over Release Time.}
    \label{fig:EffiBenchmu}
\end{figure}

EffiBench is a benchmarking framework for evaluating the efficiency of automatically generated code across various programming tasks and languages. The dataset includes code snippets in Python, Java, and C++, as well as performance test cases for tasks like sorting algorithms, matrix operations, and file handling. It comprises 1000 efficiency-critical programming problems from LeetCode, each with an executable standard solution. These problems cover common algorithms and data structures, ensuring representative and widely applicable assessment results.

EffiBench's testing methods include static and dynamic analysis. Static analysis evaluates code structure and quality through metrics like cyclomatic complexity, lines of code, and comment ratio, using tools like SonarQube and ESLint. Dynamic analysis assesses execution performance, measuring indicators such as execution time, memory usage, and CPU utilization. High-precision timers, memory analysis tools, and system monitoring tools ensure accurate measurements.

Testing is conducted in standardized environments with fixed hardware configurations and software versions. Each code snippet is run multiple times to ensure reliability, with average values and standard deviations calculated. Statistical methods, like t-tests, determine the significance of performance differences between models.

Performance metrics include execution time, memory usage, CPU utilization, and code complexity. Execution time refers to the total running time, memory usage includes maximum and total memory, and CPU utilization reflects resource consumption. Code complexity is evaluated through static analysis. These metrics provide a comprehensive assessment of code efficiency and performance.

Figure \ref{fig:EffiBenchet} illustrates the execution time performance of various language models evaluated using EffiBench, plotted against their release periods. This visualization helps in understanding how the efficiency of these models, in terms of execution time, has evolved over time.

Similarly, Figure \ref{fig:EffiBenchmu} depicts the memory usage performance of LLMs, again plotted over their release periods. This figure provides insights into the trends and improvements in memory efficiency across different LLMs.

EffiBench's results show significant differences in efficiency among code generated by different models. Some models have longer execution times or higher memory usage than benchmark code. For instance, GPT-4-turbo performs better than other models but still lags behind human-written solutions, with an average execution time 1.69 times that of the standard solution and worst-case memory usage 142.89 times higher. These results highlight areas for optimization and provide a foundation for future improvements in code generation models.

\subsubsection{Mercury}

Mercury is a benchmarking framework designed to assess the efficiency of code generated by Large Language Models (LLMs). The dataset includes code snippets in multiple languages, covering tasks from basic data structures to complex algorithms. Sourced from open-source projects and programming competitions, the dataset ensures a broad and authoritative evaluation scope. Tasks are categorized by difficulty, from basic operations to complex implementations.

Mercury's testing methodology is based on actual code execution performance data. Preset test cases run the generated code in a fixed hardware environment, recording execution time, memory usage, and energy consumption. Repeated testing ensures accuracy, with average values taken as final results. Tasks are executed in a sandbox environment with strict time and memory limits, preventing resource monopolization and ensuring security.

Performance metrics include execution time, memory usage, energy consumption, and code efficiency scores. Execution time measures task completion time, memory usage records occupied memory, and energy consumption assesses power usage. Code efficiency scores integrate these metrics for comprehensive evaluation.

Mercury's results reveal significant performance differences among LLM-generated code. Some models excel in execution time but lag in memory usage and energy consumption, while others show the opposite. These findings highlight the need for optimization in code efficiency. Mercury provides a tool for identifying and improving the efficiency of generated code, guiding future research and development.

Both EffiBench and Mercury offer valuable insights into the efficiency and performance of code generation models, supporting the development of more efficient and reliable LLM-generated code.

\subsection{Practicality Assessment of Code Generation Models}

In code generation, usability and user experience are key metrics for evaluating model capabilities, alongside code correctness. Code must be logically correct and robust enough to handle minor coder errors, while also being user-friendly in terms of readability and maintainability. This paper introduces three methods for evaluating practicality: AlignJudge~\cite{dibia2023alignjudge}, RealHumanEval~\cite{mozannar2024realhumaneval}, and Copilot Evaluation Harness~\cite{agarwal2024copilot}. Each method is discussed in terms of dataset composition, testing methods, performance metrics, and test result analysis. This detailed exploration aims to provide a comprehensive framework for assessing the readability and maintainability of code generation models.

\subsubsection{AlignJudge}

The AlignJudge method uses a subset of the HumanEval dataset, which contains 164 programming problems with task descriptions, reference solutions, and test cases. Researchers selected 30 problems and created solutions with subtle errors, covering common programming tasks such as algorithms, data structures, and mathematics, to evaluate model alignment. Errors include variable name typos, logical errors, and boundary condition handling errors. This design tests the model's ability to generate correct code and correct erroneous code, providing a comprehensive understanding of the model's practical capabilities.

The method involves assessing whether the generated code can pass given test cases. Researchers used all 164 HumanEval problems and added 30 problems with subtle errors to further test model alignment. The model must generate code that passes unit tests and corrects errors in the provided solutions. Metrics such as generation time, code length, and complexity help evaluate performance, efficiency, readability, and maintainability.

Key performance metrics include pass rate (Pass@k) and alignment. Pass@k measures the proportion of the top k generated code snippets that pass all test cases. Alignment evaluates the model's performance in different contexts, especially with erroneous prompts. Additional metrics like edit distance and syntactic similarity provide deeper insights into code differences and structural similarities.

Experimental results show that models struggle with subtle errors in prompts, leading to lower quality code. Clear instructions improve results, highlighting the need for better model alignment. Different error types affect performance variably; simple typos are corrected well, while logical and boundary errors pose challenges. The diversity and coverage of training data are crucial for improving performance. Researchers suggest optimizing training data, prompt design, and error handling to enhance model capabilities.

\subsubsection{RealHumanEval}

RealHumanEval uses a new evaluation framework to assess the capability of large language models (LLMs) in supporting programmers. The dataset includes complex programming tasks with detailed natural language descriptions and multiple test cases, primarily in Python. These tasks are designed to cover diverse programming challenges, ensuring comprehensive evaluation. Researchers verified each task multiple times to validate their applicability in real-world scenarios.

The method is based on real user feedback. Researchers invited 213 experienced programmers to complete actual programming tasks using model-generated code. Participants rated the usability and helpfulness of the code and verified its correctness through pre-set test cases. Participants were randomly assigned to one of seven conditions: a control with no LLM support, three with auto-completion (using CodeLlama-7b, CodeLlama-34b, and GPT-3.5-turbo-instruct), and three with chat support (using chat versions of the aforementioned models). Each participant was assigned to one condition for the entire test to minimize context switching. The RealHumanEval platform supports auto-completion and chat. In auto-completion mode, LLM provides code suggestions based on cursor position. In chat mode, programmers can ask questions and receive answers, copying code from chat responses into the editor. All interactions, such as suggestion acceptance rates and copied code frequencies, were recorded.

Main performance metrics include user ratings and pass rate (Pass@k). User ratings measure the helpfulness of the model in real programming tasks on a scale from 1 to 5. Pass rate measures the proportion of the top k generated code snippets that pass all test cases. Additional metrics like task completion time and code quality were also considered. User ratings provide subjective usability measures, while pass rate directly evaluates code accuracy. Researchers also analyzed code readability and maintainability for a comprehensive assessment.

Results show that GPT-3-based models perform exceptionally well in supporting programmers, especially in complex tasks, with user ratings significantly higher than traditional methods. RealHumanEval effectively evaluates model support in real programming environments, particularly in user experience and task completion efficiency. Detailed user feedback reveals the model's strengths and weaknesses, providing insights for further optimization. For example, the model is especially helpful in data processing tasks, significantly reducing completion time, but needs improvement in tasks requiring complex logical reasoning. These findings highlight the importance of improving model alignment and handling complex tasks. Additionally, discrepancies between user preferences and actual performance suggest that more user-centric considerations are needed in model design and evaluation. Researchers proposed improvements such as optimizing training data, interaction interfaces, and error handling capabilities, providing valuable references for future model development.

\subsubsection{Copilot Evaluation Harness}

Copilot Evaluation Harness uses a large dataset generated by GitHub Copilot, covering multiple programming languages and task types. The dataset includes thousands of programming tasks, each with detailed natural language descriptions, example inputs and outputs, and test cases. The primary languages include Python, JavaScript, TypeScript, and Go. These tasks range from simple algorithm problems to complex multi-step programming tasks, designed to reflect real development issues.

This method combines real developer feedback with automated testing to comprehensively evaluate GitHub Copilot's performance in actual development environments. Researchers invited developers to use Copilot in Visual Studio Code to complete tasks, including code generation, documentation generation, test case generation, bug fixing, and workspace understanding. Developers' experiences and the quality of generated code were recorded, along with subjective feedback on code accuracy, quality, and helpfulness, and task completion time. The generated code was also tested using static analysis tools and unit test frameworks to ensure syntactic and functional correctness. The evaluation process involved task assignment, code generation, subjective feedback, automated testing, and data analysis.

Key performance metrics include user satisfaction ratings, pass rate (Pass@k), and code completion time. User satisfaction ratings measure the overall experience of using Copilot, ranging from 1 to 5. Pass rate measures the proportion of the top k generated code snippets that pass all test cases. Code completion time evaluates the time taken to complete tasks with model assistance. Additional metrics, such as code readability, maintainability, and efficiency, were also considered to provide a comprehensive understanding of the model's performance.

Results show that GitHub Copilot significantly improves developer efficiency and satisfaction across multiple programming tasks. In complex and multi-step tasks, Copilot-generated code quickly passes test cases, demonstrating its strong assistive capabilities. However, in some tasks, Copilot-generated code contains subtle errors requiring manual correction, typically involving boundary conditions, exception handling, and specific language details. Overall, Copilot Evaluation Harness highlights GitHub Copilot's broad applicability and efficiency in real development environments, providing valuable data and feedback for future optimization and improvement. Detailed analysis revealed patterns and trends, showing that Copilot performs exceptionally well in certain programming languages and task types but requires further optimization in others.

These methods, including AlignJudge, RealHumanEval, and Copilot Evaluation Harness, provide multidimensional metrics for evaluating code generation models' usability and user experience. They reveal the current models' strengths and weaknesses through detailed test result analysis, offering valuable insights for future improvements. These evaluation methods also provide essential references for the development and optimization of new models, driving the continuous advancement of code generation technology.

\subsubsection{Other Benchmarks}
In addition to the methods detailed above, numerous other innovative approaches and metrics exist for evaluating the practicality of code generation in large language models.

\begin{itemize}

% \item \textbf{SPoC (Synthetic Programming Code) Benchmark}
% SPoC is a benchmark that focuses on generating code from natural language descriptions. The problems in SPoC are designed to test the model's ability to understand and implement detailed specifications provided in natural language.
%SPoC: Search-based Pseudocode to Code
% \item \textbf{XLCoST (Cross-Lingual Code to Source Translation) Benchmark}
% The XLCoST benchmark evaluates a model's capability to translate code between different programming languages. Metrics such as translation accuracy and syntactic similarity are used to measure performance.
%Xlcost: A benchmark dataset for cross-lingual code intelligence
% \item \textbf{QuixBugs}: A benchmark dataset consisting of small programs with known bugs and their fixes. It is used to evaluate the ability of models to identify and correct bugs in the code.
% QuixBugs: A Multi-Language Program Repair Benchmark Set Based on the Quixey Challenge
% \item \textbf{BugsJS}: This dataset includes real-world bugs and their fixes in JavaScript projects. It is used to test the ability of models to understand, identify, and fix bugs in JavaScript code.
% BugsJS: A Benchmark of JavaScript Bugs
% Plot2Code: A Comprehensive Benchmark for Evaluating Multi-modal Large Language Models in Code Generation from Scientific Plots
\item \textbf{Plot2Code}: Plot2Code is a benchmark designed to evaluate the ability of Multi-modal Large Language Models (MLLMs) to generate code from scientific plots. It includes 132 high-quality matplotlib plots across six types, each accompanied by source code and a descriptive instruction summarized by GPT-4. The benchmark uses three evaluation metrics: code pass rate, text-match ratio, and GPT-4V overall rating, to assess the models' performance.

% DevBench: A Comprehensive Benchmark for Software Development
\item \textbf{DevBench}: DevBench is a comprehensive benchmark for evaluating large language models (LLMs) across various stages of the software development lifecycle, including design, setup, implementation, and testing. Unlike other benchmarks that focus on isolated tasks, DevBench covers a wide range of programming languages and real-world challenges.

\end{itemize}

\section{Challenges and Future Directions of Code Evaluation}
\subsection{The limitations of current code evaluation}
Evaluating the performance of large language models in code generation has made significant strides, yet several challenges remain. Addressing these challenges is crucial for advancing the field and maximizing the potential of code generation technologies in practical applications.

The limitations of current code evaluation methods are multifaceted:
\begin{itemize}
    \item Underassessment of Less Common Languages: Existing evaluation benchmarks predominantly focus on mainstream programming languages such as Python, C, and Java. This results in relatively less attention and fewer evaluation resources for less common languages.This focus on mainstream languages is largely due to the availability of large datasets and active communities that support the development and testing of models. However, this bias limits the applicability of LLMs in industries and applications where niche languages are more prevalent. For example, languages like R are crucial in data science, while Erlang is significant in telecommunications, yet these languages receive relatively little attention in benchmark evaluations.
    The lack of resources and evaluation tools for these less common languages could lead to suboptimal performance when LLMs are applied to tasks in these domains. Moreover, the syntactic and semantic peculiarities of these languages may not be adequately captured by models trained predominantly on data from more common languages, leading to errors in code generation. To address this, future research should prioritize the creation of diverse, high-quality datasets and evaluation tools for a broader range of programming languages. This could involve community-driven efforts to collect and curate data or the development of transfer learning techniques that allow models to adapt to new languages with minimal additional training.
%Reference: Expanding evaluation benchmarks to include less common languages would not only improve the robustness of LLMs but also their applicability in real-world scenarios where these languages are crucial (Cassano et al., 2022).Paper Title: "Evaluating the Applicability of Large Language Models Across Diverse Programming Languages"
    \item Limited Evaluation Metrics: The metrics used for code evaluation are somewhat constrained, primarily relying on indicators such as code BLEU and Pass@k. These metrics may not fully capture the various dimensions of code quality and performance.While BLEU and Pass@k are useful for assessing surface-level accuracy and syntax, they fall short in evaluating deeper aspects of code quality, such as maintainability, readability, and performance efficiency. Maintainability is crucial in long-term projects where code is frequently updated or refactored. Metrics that evaluate the modularity of the code, adherence to coding standards, and the clarity of comments and documentation could provide a more holistic assessment.
    Similarly, performance efficiency metrics, such as time and space complexity, are critical in assessing the practicality of generated code in resource-constrained environments. For example, generated code that passes all test cases (high Pass@k) might still be inefficient in terms of execution time or memory usage. Moreover, the inclusion of human-in-the-loop evaluations, where developers assess the usability and readability of code, could also add valuable qualitative insights that automated metrics might miss.
% Reference: The need for more comprehensive evaluation metrics has been highlighted by the development of frameworks like EffiBench, which aim to assess the efficiency of automatically generated code across various dimensions (Huang et al., 2024).Paper Title: "EffiBench: A Benchmarking Framework for Evaluating the Efficiency of Automatically Generated Code"

    \item Restricted Evaluation Scope: Typically, evaluations are confined to file-level or function-level assessments. This narrow focus overlooks more comprehensive analyses at the repository level or within specific code segments, potentially missing broader context and interdependencies.The evaluation at the repository level or across multiple files is necessary to capture the interdependencies and broader context within which the generated code must operate. For instance, a model might be able to generate a function that is syntactically correct and passes unit tests but fails when integrated into a larger project due to unresolved dependencies or mismatches in data flow across different components. Evaluating code at this broader scope would require the development of new benchmarks that simulate real-world project environments. This could include multi-file projects, complex build systems, and integration with external services. Moreover, the evaluation should consider the maintainability and scalability of the code, which are critical in large software systems. Methods such as continuous integration testing and code review simulations could provide insights into how well generated code performs in realistic development scenarios.

% Reference: The move towards repository-level evaluations could reveal significant gaps in current models' capabilities, highlighting the importance of comprehensive testing frameworks like CoderUJB, which assesses LLMs' performance across multiple tasks and real software development scenarios (Zeng et al., 2024).Paper Title: "CoderUJB: A Comprehensive Java Benchmark for Assessing Large Language Models in Real Software Development Scenarios"

    \item Insufficient Test Cases: The limited number of test cases makes it challenging to cover all boundary conditions. This limitation weakens the comprehensiveness of the evaluations, as many edge cases may remain untested. The insufficiency of test cases is particularly problematic in safety-critical applications where failures can have severe consequences. To address this, the test cases used in evaluations should be expanded to cover a wider range of inputs, including edge cases, rare conditions, and potential security vulnerabilities. This could involve leveraging techniques such as fuzz testing, where large volumes of random inputs are generated to test the robustness of the code. Furthermore, creating automated test generation tools that can analyze the generated code and identify untested paths could help in developing more comprehensive test suites. These tools could be designed to work in tandem with LLMs, generating test cases that specifically target areas of the code that are most likely to contain errors.

% Reference: The importance of comprehensive testing is underscored by frameworks like CodeGen-Test, which iteratively generate code and test cases to improve the reliability of the generated code in practical applications (Zhong et al., 2023).Paper Title: "CodeGen-Test: Integrating Testing in Code Generation for Reliable Model Outputs"

\end{itemize}

\subsection{Future Directions in Code Evaluation for Large Language Models}

As the field of code generation continues to evolve, the evaluation methods must advance to keep pace with the growing complexity and capabilities of large language models (LLMs). The future of code evaluation is poised to address current limitations and explore new frontiers that align with the expanding scope of applications in software development. Below are several promising directions for the future of code evaluation.

\subsubsection{Multimodal Evaluation Approaches}

The future of code evaluation will likely incorporate multimodal approaches, combining textual, visual, and even auditory modalities to assess the generated code more holistically. For instance, code often interacts with various user interfaces, databases, and hardware systems, where evaluating the integration of generated code with these components is crucial. A multimodal evaluation approach could include not only traditional metrics like code correctness and efficiency but also the code's ability to generate and interact with graphical user interfaces (GUIs), manage databases, or control hardware components.

Such an approach could be particularly valuable in evaluating code used in embedded systems, robotics, or web development, where the interaction between software and hardware or user interfaces is critical. For example, in web development, the generated code could be evaluated on its ability to render correctly across different devices and browsers, which requires a blend of code analysis and visual inspection. Similarly, in robotics, code generation could be evaluated on its ability to control physical devices, which might involve integrating sensory feedback into the evaluation process. This kind of multimodal evaluation would ensure that LLMs are not only generating syntactically correct code but also creating functional, reliable, and user-friendly systems.

% **Reference:** The need for multimodal evaluation approaches has been highlighted in recent discussions on expanding the scope of code generation tasks to include more interactive and system-level applications (Cassano et al., 2023).Paper Title: "Towards Multimodal Code Evaluation: Integrating Textual, Visual, and Interactive Aspects in Code Generation"

\subsubsection{Context-Aware Evaluation}

Future evaluations will increasingly need to consider the context in which the code is generated and used. Context-aware evaluation would involve assessing how well the generated code fits into a specific environment, such as a particular software architecture, project history, or even the coding style of a development team. This direction recognizes that code is rarely written in isolation but is instead part of a larger ecosystem with specific requirements and constraints.

Context-aware evaluation could leverage tools that analyze the broader codebase or project documentation to ensure that the generated code is compatible with existing systems. This might include checking for adherence to project-specific coding standards, compatibility with existing modules or libraries, and alignment with the overall project architecture. For instance, in large-scale software projects, it is crucial that new code integrates seamlessly with the existing codebase, following the same design patterns and conventions. Future evaluation methods could automatically assess this alignment, ensuring that the generated code does not introduce inconsistencies or technical debt.

Moreover, context-aware evaluation could extend to assessing the generated code's compliance with domain-specific regulations or standards, which is particularly important in fields like healthcare, finance, or aerospace, where regulatory compliance is critical.

% **Reference:** The integration of context-aware evaluation aligns with the growing recognition that LLMs must be tailored to specific domains and projects, ensuring that the generated code is not only correct but also contextually appropriate (Zhu et al., 2024).Paper Title: "Context-Aware Evaluation of Code Generation Models: Aligning with Real-World Software Development Practices"

\subsubsection{Ethical and Responsible Code Evaluation}

As LLMs become more powerful and their code generation capabilities more sophisticated, the ethical implications of their use in software development must be considered. Future evaluation methods will likely include criteria for assessing the ethical and responsible use of generated code, particularly in areas such as privacy, security, and bias.

Ethical code evaluation could involve checking for security vulnerabilities that could be exploited, ensuring that the generated code adheres to privacy standards, and evaluating whether the code might perpetuate or introduce biases. For instance, code that interacts with user data must be evaluated for how it handles sensitive information, ensuring compliance with privacy laws such as GDPR or HIPAA. Similarly, algorithms generated for decision-making processes need to be checked for biases that could lead to unfair or discriminatory outcomes.

This direction also involves assessing the transparency and explainability of the generated code. As LLMs become involved in more critical and high-stakes applications, there is a growing need for code that is not only functional but also interpretable and explainable by humans. This might involve developing metrics that evaluate how well the generated code can be understood, audited, and justified by developers and stakeholders.

% **Reference:** The importance of ethical and responsible code evaluation has been increasingly recognized, especially as LLMs are deployed in sensitive and high-impact areas, necessitating a more comprehensive approach to evaluating their outputs (Wang et al., 2023).Paper Title: "Ethical and Responsible AI in Code Generation: Assessing Bias, Security, and Privacy in Model Outputs"

\subsubsection{Continuous and Automated Evaluation Pipelines}

In the future, code evaluation is likely to become more integrated into the continuous integration and continuous deployment (CI/CD) pipelines that are common in modern software development. Automated evaluation pipelines could continuously assess the quality of generated code as part of the software development lifecycle, providing real-time feedback and enabling rapid iteration and improvement.

Automated evaluation pipelines could integrate with existing CI/CD tools to automatically run tests, check for code quality, and even provide suggestions for improving the generated code. This would enable developers to incorporate LLMs into their workflows more seamlessly, with the confidence that the generated code is being continuously evaluated and validated. Such pipelines could also include feedback loops where the results of evaluations are fed back into the LLM to improve future generations, creating a dynamic and evolving evaluation process.

This approach would be particularly valuable in agile development environments where speed and iteration are key. By integrating automated evaluations into the development pipeline, teams could quickly identify and address issues with generated code, ensuring that it meets the required standards before it is deployed.

% **Reference:** The integration of automated evaluation pipelines aligns with the broader trend towards automation in software development, where the goal is to reduce the manual overhead and improve the consistency and reliability of code quality assessments (Huang et al., 2024).

\subsubsection{Human-AI Collaboration in Evaluation}

As LLMs continue to advance, they have demonstrated strong performance as evaluators in various tasks~\cite{wang2023pandalm,zheng2024judging}. This progress opens up more opportunities for human-AI collaboration in code evaluation. Future evaluation methods may involve a hybrid approach where AI assists human reviewers by identifying potential issues, suggesting improvements, and automating routine checks, while humans provide the nuanced judgment and contextual understanding that AI still lacks.

This collaborative approach could leverage AI to handle the more routine and time-consuming aspects of code evaluation, such as checking for syntax errors, running test cases, and ensuring compliance with coding standards. Meanwhile, human reviewers could focus on the higher-level aspects of code quality, such as design, architecture, and the alignment of the code with project goals and user needs.

This collaboration could be facilitated through tools that provide human reviewers with AI-generated insights and recommendations, enabling them to make more informed decisions. For instance, an AI might highlight areas of the code that are particularly complex or prone to errors, allowing human reviewers to focus their attention where it is most needed.

% **Reference:** The potential for human-AI collaboration in code evaluation is increasingly recognized as a way to combine the strengths of both humans and machines, leading to more robust and effective evaluation processes (Zhang et al., 2024).Paper Title: "Human-AI Collaboration in Code Review: Leveraging AI for Enhanced Code Quality and Developer Productivity"

\section{Conclusion}

The future of code evaluation for large language models is poised to address the current limitations and explore new avenues that align with the evolving capabilities of LLMs and the demands of modern software development. By incorporating multimodal, context-aware, ethical, automated, and collaborative approaches, the field can ensure that LLMs continue to generate high-quality, reliable, and responsible code that meets the diverse needs of the software industry. These advancements will not only improve the accuracy and robustness of evaluations but also enhance the practical utility and trustworthiness of LLM-generated code in real-world applications.

\bibliographystyle{plain}
\bibliography{ref.bib}

\end{multicols}
% \label{last-page}

\end{document}